\documentclass[11pt]{article}
\usepackage{geometry} 
\usepackage[parfill]{parskip}    
\usepackage{graphicx}
\usepackage{amssymb}
\usepackage{epstopdf}
\title{Exact Solutions for Modified Burgers Vortex}
\author{David Rollins}

\begin{document}
\maketitle

\section{Introduction}
The vortex stretching process is relevant to energy transport at various scales in turbulent flow  and in vortex reconnection. For these, reasons, exact solutions for vortex flow are useful. Robinson and Saffman [1] examined the Burgers vortex solution and found closed form steady solutions for special parameter values. Unsteady 2D Burgers vortex solutions have been used to model the spanwise structure of turbulent mixing layers (Lin and Corcos [2], Neu [3]). Unsteady axisymmetric Burgers vortex solutions are used in modeling the fine-scale structure of homogeneous incompressible turbulence (Townsend [4], Lundgren [5]).

Shivamoggi [6] considered a modified Burgers vortex that describes convection of vortex lines toward the $z$-axis and stretching along the $y$-axis. The straining flow is externally imposed so vorticity is decoupled from this flow. Exact solutions were found in the special case where the flow parameter is constant. In this paper, new exact solutions are demonstrated for a particular non-constant case.

\section{Modified 2D Burgers Vortex}

Consider a modified Burgers vortex with the velocity field given by
\begin{equation}
\label{velfield}
\mathbf{v}=(-\gamma(t)x,\gamma(t)y, W(x,t)).
\end{equation}
(\ref{velfield}) describes convection of the vortex lines toward the $y$-axis and stretching along the $y$-axis. This velocity field has streamlines that are the same in each plane parallel to the $x$-$y$ plane. Note that the flow associated with the vortex is perpendicular to the plane of the uniform straining flow in distinction to that of the standard Burgers vortex flow. This suggests that this flow can be used to model mixing layer flow or jet flow. The vorticity corresponding to (\ref{velfield}) is given by
\begin{equation}
\label{vort}
\mathbf{\omega}=\triangledown \times \mathbf{v}= (0, -\partial W/\partial x,0)
\end{equation}
so that vortex lines are aligned along the $y$-axis as for the standard Burger's model.

Using (\ref{velfield}) and  (\ref{vort}), the vorticity conservation equation
\begin{displaymath}
\frac{\partial\omega}{\partial t}+(\mathbf{ v}\cdot \triangledown)\omega = (\omega \cdot \triangledown)\mathbf{v} + \nu \triangledown^{2} \omega
\end{displaymath}
becomes
\begin{equation} \label{vortcon}
\frac{\partial\Omega}{\partial t}-\gamma x \frac{\partial\Omega}{\partial x}=\gamma \Omega + \nu\frac{\partial^{2}\Omega}{\partial x^{2}}
\end{equation}
where $\nu$ is the kinematic viscosity and $\Omega$ is the vorticity component
\begin{displaymath}
\Omega \equiv \frac{\partial W}{\partial x}.
\end{displaymath}

The introduction of the dimensionless independent variables
\begin{equation}
\label{dimvar}
\xi =\sqrt{\frac{\gamma(t)}{\nu}}, \quad \tau = \int_{0}^{t} \gamma(t') \, dt'
\end{equation}
results in (\ref{vortcon}) becoming
\begin{equation}
\label{vortcon2}
\frac{\partial\Omega}{\partial \tau}+\frac{\dot{\gamma}}{2\gamma^{2}} \xi \frac{\partial\Omega}{\partial \xi}= \frac{\partial}{\partial \xi}(\xi\Omega)+\frac{\partial^{2}\Omega}{\partial \xi^{2}}.
\end{equation}

Note that the coefficient of the second term on the left-hand side of (\ref{vortcon2}) is $\tau$ dependent through the defining relation between $\tau$ and $t$ in (\ref{dimvar}). In general exact solutions are not possible except in special cases. Shivamoggi [6] considered the case $\dot{\gamma}=0$, or $\gamma=$ constant, which can be solved exactly. Another case where an exact solution is possible is when
\begin{displaymath}
 \dot{\gamma}=2c_{1} \gamma^{2}
\end{displaymath}
which has solution
\begin{displaymath}
\gamma(t)=\frac{-1}{2c_{1}t+c_{2}}
\end{displaymath}
where $c_{1}$ and $c_{2}$ are constants chosen so that $\gamma(t)>0$. This guarantees that the coordinates defined in (\ref{dimvar}) are real valued. Note that $\gamma=$ constant is a special case when $c_{1}=0$ and Shivamoggi's solution [6] is included. Equation (\ref{vortcon2}) is now
\begin{equation}
\label{vortcon3}
\frac{\partial\Omega}{\partial \tau}=\Omega+ \alpha \xi \frac{\partial\Omega}{\partial \xi}+\frac{\partial^{2}\Omega}{\partial \xi^{2}}.
\end{equation}
where $\alpha=1-2c_{1}>$. The boundary conditions are
\begin{displaymath}
|\xi| \rightarrow \infty : \quad \Omega \rightarrow 0.
\end{displaymath}

\section{Exact Solutions}
If it is assumed that $\partial\Omega / \partial \tau =0$ in (\ref{vortcon3}), then the exact solution is given in terms of the parabolic cylinder function:
\begin{displaymath}
\Omega (\xi)=C_{1} e^{-\alpha\xi^{2}/4}D_{1/\alpha -1}(\xi)
\end{displaymath}
and the $z$-velocity component is
\begin{displaymath}
W(\xi)=C_{1}\sqrt{\frac{\gamma(t)}{\nu}}\int_{0}^{\xi} e^{-\alpha\eta^{2}/4}D_{1/\alpha -1}(\eta)\, d\eta.
\end{displaymath}
where $C_{1}$ is an arbitrary constant.

Separable solutions are also possible. Assuming solutions of the form
\begin{displaymath}
\Omega(\xi, \tau)=h(\xi)e^{-\lambda \tau}
\end{displaymath}
where $\lambda$ is a constant, the from (\ref{vortcon3}), $h$ satisfies
\begin{equation}
\label{eigen}
\frac{d^{2} h}{d \xi^{2}}+\alpha\xi \frac{d h}{d \xi}+h=-\lambda h.
\end{equation}

For bounded solutions of (\ref{eigen}) to exist, it is required that
\begin{displaymath}
\lambda_{n}=(n+1)\alpha-1
\end{displaymath}
which gives the solution
\begin{displaymath}
h_{n} (\xi)=(-1)^{n} e^{-\alpha\xi^{2}/4}H_{n}(\sqrt{\frac{\alpha}{2}}\xi)
\end{displaymath}
with $n=0,1,2,...$ and $H_{n}$ the Hermite polynomials.

\section{Discussion}
Exact solutions have been found for a particular type of time dependence in a modified Burgers vortex flow. These solutions reduce to the special case in [6] when $\alpha=1$.

\section{References}
\begin{enumerate}
  \item A.C. Robinson, P.G. Saffman, Stud. Appl. Math. \textbf{70}, 163, (1984).
  \item S.J. Lin, G.M. Coros, J. Fluid Mech. \textbf{141}, 139, (1984).
  \item J. Neu, J. Fluid Mech. \textbf{143}, 253, (1984).
  \item A.A. Townsend, Proc. Roy. Soc. (London) A \textbf{208}, 5343, (1951).
  \item T.S. Lundgren, Phys. Fluids \textbf{25}, 2193, (1982).
  \item B.K. Shivamoggi, Eur. Phys. J. B \textbf{49}, 483, (2006).
\end{enumerate}

 \end{document}